# Lattice Investigations of the Running Coupling

P. Weisz[a]

[a]Max-Planck-Institut für Physik, Föhringer Ring 6, D-80805 München, Germany

A review of investigations of running couplings using lattice techniques is given. This includes i) studies of the running of particular non-perturbatively defined renormalized couplings in pure gauge theories over a range of energies, and ii) how estimates of $\alpha_{\overline{\rm MS}}(m_Z)$ in lattice QCD are presently obtained.

## 1. INTRODUCTION

A widely accepted property of QCD is asymptotic freedom (AF), which allows the computation of certain aspects of high energy $E$ processes as a perturbative series in a coupling which depends on $E$ and goes to zero as $E \to \infty$. There are infinitely many physically admissible non-perturbative definitions of running couplings. At low energies (LE) these can have completely different behaviors from one another, but at high energy (HE) they behave similarly and can be related to one another using perturbation theory (PT).

The currently most widely used scheme for the analysis of experimental data in HE physics is the $\overline{\rm MS}$ scheme [1] using dimensional regularization. However it is important to bear in mind that $\alpha_{\overline{\rm MS}}$ is only operationally defined in PT. Its extraction from experimental (or numerical) data is not entirely straightforward. Obviously it requires the dropping of higher order terms which have not been computed. But more awkward is the fact that the nature of the perturbative series is unknown and there are principle uncertainties [2] due to the neglect of non-perturbative effects; these are typically $O(\Lambda_{\overline{\rm MS}}^2/E^2)$ (where $\Lambda_{\overline{\rm MS}}$, the so-called lambda-parameter, is a characteristic hadronic scale), only a priori negligible at really high energies. For example to take the uncertainty of such effects into account it has been suggested [3] that the error in $\alpha_{\overline{\rm MS}}(m_\tau)$ now quoted $0.36(3)$ should be doubled (yielding $0.122(7)$ for $\alpha_{\overline{\rm MS}}(m_z)$ instead of $0.122(4)$).

Values of $\alpha_{\overline{\rm MS}}$ extracted from various experiments in an energy range up to the Z-peak are qualitatively consistent with the running predicted by the renormalization group (RG) equations using perturbative $\beta$-functions, but the statistical and systematic error bars are too big to make definite statements. A compilation of values for $\alpha_{\overline{\rm MS}}(m_Z)$ from various experiments was made last year by Webber [4]. There is tendency for the central value extracted from low energy (and run to $m_Z$ using RG) to be slightly lower than that obtained directly from the LEP collider at the Z-peak. An often posed question in present phenomenology [5] is whether this is due to underestimating the systematic errors referred to above or whether the potential discrepancy should be taken seriously and interpreted as a manifestation of new physics.

A definition of QCD requires a non-perturbative regularization e.g. the lattice. The possibility to compute running couplings in asymptotically free theories through numerical simulations of the corresponding lattice theories has recently attracted much attention [6]-[33]. There are two main directions of research: in the first one actually *studies the running* of a coupling over a range of energies; in the second one uses certain measurements (see sect.6) to estimate $\alpha_{\overline{\rm MS}}$ at relatively low energies and then reaches a value of $\alpha_{\overline{\rm MS}}(m_Z)$ *assuming applicability of perturbative RG running*.

The definitions of the coupling which have been investigated on the lattice are given in sect.2. If these are measurable to sufficient accuracy, one can actually test if the behavior is really that predicted by the RG with the perturbative $\beta$-functions. There is no rigorous proof that this must be the case, and some authors [34] favor the scenario that although the continuum limit of lattice QCD may exist and describe the LE physics properly, the HE behavior will not be that described by AF (i.e. typical couplings will not approach zero at HE). Such scenarios can practically not be distinguished experimentally, but the question can be addressed on the lattice.

Taking the conventional scenario as a working hypothesis, the determination of $\alpha_{\overline{\rm MS}}(m_Z)$ with a reliable error estimate requires various basic ingredients: 1) an accurate determination of a reference energy scale (see sect.3.); 2) the introduction of a non-perturbatively defined renormalized coupling which can be measured accurately on the lattice over a wide range of energies; and 3) a careful treatment of the continuum limit. Having these one can see at which energy the expected perturbative behavior actually sets in, and thereon one can confidently run to even higher energies e.g. $m_Z$, where one converts to the $\overline{\rm MS}$-scheme. In order to avoid introducing an additional error > 1% at this final stage one needs the knowledge of the relation to $\alpha_{\overline{\rm MS}}$ to 2-loops (see sect.4).

The program above has so far only been achieved for pure gauge theories [7]-[13] (see sect.5). To obtain preliminary estimates of $\alpha_{\overline{\rm MS}}(m_Z)$ in QCD a shortcut has been proposed which involves the dropping of steps 2) and 3), and using instead a modified bare PT as explained in sect.6, to estimate $\alpha_{\overline{\rm MS}}$ at some lower energy and then using perturbative running to $m_Z$. It turns out in pure gauge theory that this procedure reproduces the results of the more detailed investigation, but one would not be able to say this a priori since *the systematic errors due to cutoff effects are not really controlled in the shortcut approach*.

In QCD the method above has been invoked [21]-[33] producing results (already quoted in the Particle Data Tables) consistent (at present) with experimental determinations. Such computations have been made in the quenched approximation and in full QCD with $n_f = 2$; these are reviewed in sect.7. In sect.8 we describe the status of the Alpha-collaboration project to study the actual running. Finally in sect.9 we briefly discuss the treatment of the heavy quark thresholds.

## 2. DEFINITIONS OF COUPLINGS

### 2.1. Couplings from the potential

Various definitions of the coupling involve the static potential $V(r)$ [14]-[19]. The first is

$$\alpha_{\rm q\bar{q}}(q) = C_{\rm F}^{-1} r^2 F(r), \quad q = 1/r, \quad (1)$$

where $F(r) = {\rm d}V(r)/{\rm d}r$ is the force (up to a sign) and $C_{\rm F} = (N^2 - 1)/2N$. Alternatively one can consider the Fourier transform $\tilde{V}(q)$ of the potential and define

$$\alpha_{\rm V}(q) = -\frac{1}{4\pi} C_{\rm F}^{-1} q^2 \tilde{V}(q). \quad (2)$$

Yet another recent suggestion [19] is to define couplings by using $R \times T$ Wilson loops with $R/T = x$ fixed instead of taking the $T \to \infty$ limit.

With these definitions it is difficult to get $\alpha(q)$ in a large momentum range, despite the fact that the quality of data on Wilson and Polyakov loops is now very good.

### 2.2. From the 3-gluon vertex

Couplings defined from the 3-point vertices were previously investigated in PT and considered (despite gauge dependence) "more physical in some respects" than $\alpha_{\overline{\rm MS}}$. Recently a project has been initiated [20] to measure a renormalized coupling based on the triple gluon vertex $G^{(3)}$ (in the Landau gauge). Again the momentum range which can be covered is very small at present.

### 2.3. Couplings running with the volume

The first coupling which runs with the box size to be considered in gauge theories, is based on the Schrödinger functional (SF) [7] $Z(L, \eta)$. This is the partition function on a finite volume $L \times L^3$ with periodic boundary conditions in the spatial directions and Dirichlet bc's in the 'time' direction. The fields at the boundaries ($x_0 = 0, L$) have to be chosen appropriately (see [7]) and depend on a parameter $\eta$. The coupling is then defined as a response of the system to changing the bc's

$$\alpha_{\rm SF}(q)^{-1} \propto \frac{\partial}{\partial \eta} \ln Z(L, \eta)|_{\eta=0}, \quad q = 1/L. \quad (3)$$

This coupling vanishes exponentially as $L \to \infty$.

A second definition is based on the ratio of correlations of Polyakov loops [11]:

$$\alpha_{\rm TP}(q) \propto \frac{\langle P_x(0)P_x(L)\rangle}{\langle P_z(0)P_z(L)\rangle}, \quad q = 1/L, \quad (4)$$

where the gauge field satisfies periodic bc's in the $z$-direction but twisted bc's in the $x, y$ directions. At low energies this coupling goes to a constant.

These couplings seem at first rather exotic, but at HE they can be matched to infinite volume physical couplings using PT. The basic advantages of couplings which run with the box size $L$ were explained in detail in ref.[9]. Firstly they can be accurately measured over a wide range of energies. Secondly, a very important aspect, is that there is a well-controlled extrapolation to the continuum limit.

### 2.4. Generalities

All the couplings above can be defined in the theories including fermions. Note, as mentioned previously, they all behave asymptotically the same at HE but very differently in the LE (nonperturbative) regime. Given one coupling one can trivially define from it another valid coupling having different LE behavior. An illuminating example was given by Michael last year [16]; for pure gauge theories consider the coupling $\tilde\alpha_{\rm F}(q) = C_{\rm F}^{-1}r^2(F(r) - K)$, where $K \equiv F(\infty)$ is the string tension. At HE $\tilde\alpha_{\rm F}(q)$ and $\alpha_{q\bar q}(q)$ have the same perturbative expansion and differ only by a non-perturbative term $\propto \exp[-1/4\pi b_0 \alpha_{q\bar q}(q)]$ [1] which becomes negligible ($< 1\%$) for $q \gg 10 {\rm GeV}$.

### 3. SETTING THE SCALE

To set the energy scale one needs a physical infinite volume observable $q_0$ (with mass dimension). At a given bare coupling $\alpha_0$, a measurement of $aq_0$ then gives the lattice spacing in units of $q_0$ (or in units ${\rm GeV}^{-1}$ if $q_0$ is assigned its physical value). The dependence on the observable will not be considerable provided one is in the scaling region where cutoff effects are small. The *working hypothesis*, motivated by the work of Symanzik

---
[1]Here and in the following $b_0, b_1$ denote the universal 1,2-loop $\beta$-function coefficients: $b_0 = \left(\frac{11}{3}N - \frac{2}{3}n_f\right)\left(\frac{1}{16\pi^2}\right)$ and $b_1 = \left(\frac{34}{3}N^2 - \left\{\frac{13}{3}N - N^{-1}\right\}n_f\right)\left(\frac{1}{16\pi^2}\right)^2$.

[35], is that physical quantities approach their continuum limit with power corrections in $a$.

The problem is to find a scale which can be measured accurately since errors $\delta q_0$ in a scale $q_0$ give errors $\delta\alpha/\alpha \sim 0.2\delta q_0/q_0$ in $\alpha$ at $m_Z$ i.e. one needs the scale to $< 5\%$ to avoid introducing an error $> 1\%$.

For the pure gauge theory one can consider for example a glueball mass or the string tension $K$. Starting from the potential $V$ a better parameter is Sommer's scale $r_0$ [36]:

$$r_0^2 F(r_0) = 1.65. \quad (5)$$

It is easier to extract from the data than $K$ and its definition extends to the theory with fermions. Actually $r_0$ is not obtained directly from experiment; the value 1.65 above is chosen since the phenomenological potential describing the spectrum of charmonium has this value for $r_0 \sim 0.5{\rm fm}$. For $N = 3, \sqrt{K}r_0 \sim 1.2$ [36].

One could contemplate eventually using some spectral value $m_\rho$ or $m_{\rm P}$. But at present this is rather difficult because of the required quark mass extrapolation $m_q \to 0$, and in the case of $m_\rho$ one needs to take into account resonance effects. Alternatively one could take meson decay constants e.g. $f_\pi$, but this involves careful non-perturbative evaluation of the axial current normalization constant.

Setting the scale with mass splittings in heavy quarkonium [23,21] $1P - 1S, 2S - 1S$ has various advantages. Firstly they empirically depend only weakly on the heavy quark mass and secondly they are expected to be rather insensitive to very light quark effects. The main problem is that in presently conceivable simulations of relativistic fermions $m_c a \sim 1$, so that cutoff effects are potentially large. The way to overcome this is to use an appropriate effective lagrangian for the heavy quark system e.g. a non-relativistic treatment [28,21] (see also sect.7).

### 4. SCHEME DEPENDENCE & BOOSTS

#### 4.1. Renormalized coupling relationships

As mentioned before, at HE different schemes are related by PT. For example through the series

$$\alpha_{\overline{\rm MS}}(sq) = \alpha_{\rm X}(q) + c_1^{\rm X}(s)\,\alpha_{\rm X}(q)^2 + c_2^{\rm X}(s)\,\alpha_{\rm X}(q)^3 + .. (6)$$

Table 1
Coefficients $c_1$ in eq.(6) for massless quarks

| Scheme X | $c_1^X(1)$ |
|---|---|
| $\overline{\text{MS}}$ | $4\pi b_0 (\ln 4\pi - \gamma_E)$ |
| $q\bar{q}$ | $(-31N + 10n_f)/36\pi$ |
| V | $c_1^{q\bar{q}}(1) - 8\pi b_0 \gamma_E$ |
| SF, $N=2, n_f=0$ | 0.94327(4) |
| SF, $N=3$ | $1.256 + 0.023 n_f$ |
| TP, $N=2, n_f=0$ | $-0.5584(4)$ |

Table 2
Values of perturbative coefficients

| Coefficient | Value |
|---|---|
| $k_1$ | 2.135730074078457(2) |
| $k_2$ | $-2.8626215972(6)$ |
| $k_3$ | 1.24911585(3) |
| $a_1$ | 0.0063537 |
| $a_2$ | $-0.0181240$ |
| $a_3$ | 0.0185223 |

one can convert from the X-scheme to the $\overline{\text{MS}}$ scheme [1]. Since $\alpha \sim 0.1$ at the Z-peak, it is necessary to know the coefficient $c_2^X$ if we want to obtain results for $\alpha_{\overline{\text{MS}}}(m_Z)$ to a precision of 1%.

The $s$-dependence of the coefficients is fixed by the RG equation:

$$c_1^X(s) = -8\pi b_0 \ln s + c_1^X(1), \qquad (7)$$

$$c_2^X(s) = c_1^X(s)^2 - 32\pi^2 b_1 \ln s + c_2^X(1) - c_1^X(1)^2. \quad (8)$$

The $c_1$ have been computed for all the schemes mentioned in sect.2 and are given in Table.1.

The reason for introducing the scale factor ("boost") $s$ in eq.(6) is the intuition that before making practical use of the series, $s$ should be fixed so that the low-order terms are reasonably small (as far as possible). One can e.g. chose the value of $s = s_0$ such that $c_1(s)$ vanishes; this is just given by the ratio of the $\Lambda$-parameters [2]

$$c_1^X(s_0^X) = 0 \text{ for } s_0^X = \exp\left\{\frac{c_1^X(1)}{8\pi b_0}\right\} = \frac{\Lambda_{\overline{\text{MS}}}}{\Lambda_X}. \quad (9)$$

For the special case of the $\overline{\text{MS}}$-scheme we have

$$\alpha_{\overline{\text{MS}}}(s_0^{\text{MS}} q) = \alpha_{\text{MS}}(q) \qquad (10)$$

to all orders.

Apart from this rather trivial example, at present the only known 2-loop parameter $c_2$ is for the scheme $X = \text{SF}$ for $N=2, n_f=0$ [37,38].

### 4.2. Expansions in the bare parameter

To get the relationships between the different schemes it is often easier (albeit not necessary) to first expand each coupling in powers of the bare coupling $\alpha_0 = g_0^2/4\pi$,

$$\alpha_X(q) = \alpha_0 + d_1^X(qa)\,\alpha_0^2 + d_2^X(qa)\,\alpha_0^3 + \ldots \quad (11)$$

In the continuum limit the bare coupling and the lattice spacing $a$ are simultaneously taken to zero in such a way that $a$ is an exponentially vanishing function of $\alpha_0$. In this context it is consistent to drop all terms of order $a^p, p > 0$ contributing to the coefficients $d_n$, which then reduce to polynomials in $\ln(qa)$ as determined by the RG equation cf eqs.(7,8). When we combine the relations for different schemes to eliminate $\alpha_0$ the dependence on $a$ also drops out.

The coefficients $d_n^X(1)$ depend of course on the lattice action used. Here we consider only a 4-dimensional hypercubic lattice and for the pure gauge SU($N$) theory the standard Wilson action. The coefficients $d_1^X$ for all schemes can be obtained from the knowledge of $c_1^X$ and $d_1^{\overline{\text{MS}}}$ which was computed nearly 15 years ago by A. and P. Hasenfratz [39]:

$$d_1^X(1) = d_1^{\overline{\text{MS}}} - c_1^X(1), \qquad (12)$$

$$d_1^{\overline{\text{MS}}}(1) = -\frac{\pi}{2N} + k_1 N, \qquad (13)$$

with $k_1$ given in Table 2.

The first 2-loop coefficient $d_2$ to be obtained was for the SF-scheme for the case of SU(2), by Narayanan and Wolff [37]

$$d_2^{\text{SF}}(1) - d_1^{\text{SF}}(1)^2 = 2.5375(4). \qquad (14)$$

An estimate for $d_2^{\overline{\text{MS}}}(1)$ has been obtained by Ellis and Martinelli [40,41] on the basis of "tadpole dominance" (as first proposed by Parisi [42]).

---
[2] Note that the value of $c_2(s_0^X)$, if positive, is (at least for $N=2,3, n_f=0$) very close to the minimum value of $c_2$, since $c_2^X(s) \geq c_2^X(s_0^X) - (2\pi b_1/b_0)^2$ for all $s$.

But a complete computation was only presented recently by Lüscher and myself [38]. Due to the complications of lattice PT we decided to use the background field technique [43,44], which has previously been applied to compute the 1-loop coefficient $d_1^{\overline{\rm MS}}(s)$ [45].

In ref.[46] the background field method on the lattice is described: this discussion leans heavily on the work of Kluberg-Stern and Zuber [44] in the continuum, and on the proof of renormalizability of lattice PT by Reisz [47]. The application to the computation of $d_2^{\overline{\rm MS}}$ was given in [38]. For the evaluation of lattice Feynman integrals in the continuum limit, we employed a newly developed position space technique [48]. A crucial ingredient of this is an observation of Vohwinkel that the free boson (lattice) propagator can be computed recursively algebraically in terms of its values at the corners of the unit cube at the origin.

To determine $d_2^{\overline{\rm MS}}(1)$ the background field propagator must be worked out to 2-loops on the lattice, and matched with the corresponding quantity in the continuum theory using the $\overline{\rm MS}$ scheme of dimensional regularization which has been previously computed by Ellis [41] [3].

The final result is

$$d_2^{\overline{\rm MS}}(1) - d_1^{\overline{\rm MS}}(1)^2 = \frac{3\pi^2}{8N^2} + k_2 + k_3 N^2, \qquad (15)$$

where $k_2, k_3$ are given in Table 2.

### 4.3. The coefficients for the Plaquette

The (normalized) plaquette expectation value $P$ [49] is now known to 3-loops [50]:

$$P = 1 - p_1 \alpha_0 - p_2 \alpha_0^2 - p_3 \alpha_0^3 + \ldots, \qquad (16)$$

with

$$p_1 = C_{\rm F} \pi, \qquad (17)$$

$$p_2 = -C_{\rm F} N \left( 2k_2 + \frac{5\pi^2}{12} + \frac{\pi^2}{4N^2} \right), \qquad (18)$$

$$p_3 = C_{\rm F} \frac{N^2}{4} \left( a_1 + a_2 N^{-2} + a_3 N^{-4} \right) (4\pi)^3, \qquad (19)$$

---
[3] and checked by A. van de Ven and M. Lüscher.

with $a_n$ given in Table 2, and $k_2$ is the same constant as in eq.(15). The coefficients $a_n$ have been checked using MC methods [51]. These computations (and those of the previous subsection) illustrate the rapid growth of the application of algebraic and numeric computer methods for the evaluation of lattice Feynman diagrams.

## 5. INVESTIGATING THE RUNNING

### 5.1. The SF- and TP-couplings

Last year Guagnelli [13] presented the results of the Alpha-collaboration study of the universality of the continuum limit in pure SU(2) Yang-Mills theory. The two couplings $\alpha_{\rm SF}$ and $\alpha_{\rm TP}$ (see eqs.(3,4)) were accurately determined [12] over a large range of momenta given in units of Sommer's scale $r_0$ [36]. A crucial feature to appreciate in these analyses is that the results are extrapolated to the continuum limit, using a multi-lattice approach. The method has been discussed in detail in many places (see e.g.[10]). The idea is that a sequence of lattices with parameters $(\beta = \beta_i, L/a = L_i/a)$ are determined such that say the coupling $\alpha_{\rm SF}(L)$ remains fixed, and then for the same parameters $\alpha_{\rm SF}(2L)$ is measured and the resulting sequence is extrapolated to the continuum limit. Here, (as referred to in sect.3), the working hypothesis is that cutoff effects fall as powers in $a/L$. It was found that the data was certainly consistent with such a behavior. The procedure is then repeated for $\alpha_{\rm SF}(L)$ fixed at the previously extrapolated value of $\alpha_{\rm SF}(2L)$, in order to obtain the running.

For a sequence of lattices with fixed value of $\alpha_{\rm SF}(L)$ we also measured the corresponding values of $\alpha_{\rm TP}(L)$. These approached a definite value (again in the expected way) providing additional evidence for the existence of the continuum limit.

A remarkable property of both couplings (in the continuum limit) is that their evolution in the momentum range covered by the numerical simulations is well described by the 2-loop approximation to the Callan-Symanzik $\beta$–function [8,12] [4],

---
[4] A perfect fit of the SF-data was in fact obtained by adding an effective 3-loop term with coefficient $b_2^{\rm eff} = 0.35(12)/(4\pi)^3$. With $c_2^{\rm SF}$ and the known value of $b_2^{\overline{\rm MS}}$ [52] we may now calculate the exact coefficient $b_2^{\rm SF} =$

only $\alpha_{\rm TP}$ at the lowest energy shows a slight deviation. We also found that the perturbative relation between the couplings (unfortunately only known presently to 1-loop) works better if one chooses the corresponding boost $s$ (cf sect.4) to be equal to the appropriate ratio of $\Lambda$-parameters rather than just setting $s = 1$. These features can be seen in Fig.1.

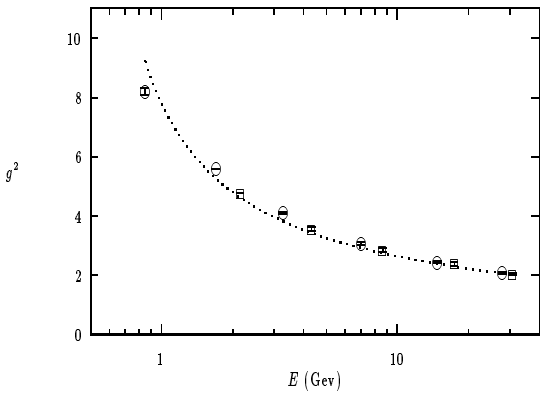

Figure 1. A comparison of $\bar{g}_{\rm TP}^2(E)$ (circles) and $\bar{g}_{\rm SF}^2(\Lambda_{\rm SF} E/\Lambda_{\rm TP})$ (boxes). The dotted line is 2-loop evolution.

Since $\alpha_{\rm SF}$ runs with the expected RG behavior in the HE range measured, it is reasonable to continue the integration up to even higher energies and there convert to the $\overline{\rm MS}$-scheme. If we choose $s = \Lambda_{\overline{\rm MS}}/\Lambda_{\rm SF}$, (which proved successful above), then at $q = 200/r_0$ (which corresponds to about 80 GeV in physical units), the result using the formulae only to 1-loop is [12] (SU(2))

$$\alpha_{\overline{\rm MS}}(200/r_0) = 0.1288(15)(21), \quad (20)$$

where first error is the total statistical error, while the second is $\sim \alpha_{\rm SF}^3$ (indicating the expected order of magnitude of the 2-loop correction). With the 2-loop coefficients eqs.(14,15) at hand we can now do better. In fact one finds a very small 2-loop coefficient $c_2^{\rm SF}(s_0^{\rm SF}) = 0.0576(5)$, which has practically no influence on the central value of

---

0.1797(3)/$(4\pi)^3$. There is no reason to expect that $b_2^{\rm SF}$ is equal to $b_2^{\rm eff}$, but it is reassuring that both coefficients have the same sign and order of magnitude.

Table 3
Evaluations of $\alpha_{\overline{\rm MS}}(37/r_0)$ for SU(3), $n_f = 0$.

| Method | $\alpha_{\overline{\rm MS}}(37/r_0)$ |
|---|---|
| SF,1-lp,$s=1$ | 0.1108(23)(14) |
| SF,1-lp,$s=s_0^{\rm SF}$ | 0.1141(24)(14) |
| $\alpha_{q\bar{q}}$,$\beta = 6.5$;2-lp ev. | 0.1180(21) ($^{+0}_{-60}$) |
| eq.(28),$\beta = 6.5$;1-lp | 0.1121(14) |
| eq.(28),$\beta = 6.5$;2-lp | 0.1134(2) |
| eqs.(29-31),$\beta = 6.0$ | 0.1113(11) |
| eqs.(29-31),$\beta = 6.4, 6.5$ | 0.1126(10) |

$\alpha_{\overline{\rm MS}}(q)$ and the result eq.(20) is modified to

$$\alpha_{\overline{\rm MS}}(200/r_0) = 0.1289(15)(3). \quad (21)$$

The main impact of the 2-loop correction is that the estimated error from the neglected higher order terms (now $\sim \alpha_{\rm SF}^4$) has been reduced from a few percent to a fraction of a percent and is now less than the statistical error.

The SF-coupling has also been measured in the SU(3) theory [10]. Again the expected running was observed and the results are qualitatively similar to the case of SU(2). Unfortunately the result for $d_2^{\rm SF}$ for $N = 3$ is not yet available [5], so formulae are presently restricted to 1-loop. The result for $\alpha_{\overline{\rm MS}}$ from the data point at highest energy is given in Table 3 for two choices of the boost $s$.

### 5.2. Couplings from the potential

The data on Wilson loop expectation values and on their correlations now available is of very high quality, both in SU(2) and SU(3) Yang-Mills theory, thanks largely to the Wuppertal group [17] and UKQCD [14,15], so that one now has information on the force and on the shape of the flux tube [53], over a range of distances. If the force is defined as in eqs.(2.5-2.7) of ref.[36] then the cutoff effects seem to be rather small. The behavior of the resulting running coupling $\alpha_{q\bar{q}}$ at short distances is consistent with the expected perturbative RG behavior. Instead of showing $\alpha_{q\bar{q}}$, I would here like to show Fig.2, the force itself as a function of distance in units of $r_0$, which emphasizes the non-perturbative aspects. Fig.2 was produced by Lüscher from data (in Table 3 of ref.[36]) for $\beta$-values 2.5-2.7. The dotted line is the curve $(r^{-2} - r_0^{-2})\pi/12 + 1.65 r_0^{-2}$. One indeed observes remarkable evidence for the presence of Lüscher's universal $\pi/(12r^2)$ force [54]! Also it is rather surprising that this formula, which is an asymptotic formula for the non-perturbative large $r$ regime, actually describes the data quite well up to quite HE - yet another example of precociousness (but this time from the other end).

---

[5] the computation is under way by A. Bode

Figure 2. Force in SU(2) pure gauge theory.

The force in the SU(3) [55] theory looks very much the same as in Fig.2; the corresponding plot (updating that in ref.[55]) can be obtained from Wittig. As a typical result for $\alpha_{\overline{MS}}$, take the data for $\alpha_{q\bar{q}}(q)$ at $\beta = 6.5$, which covers the range up to $q \sim 7/r_0$. Then using the 1-loop perturbative relation (with $s = 1$) to go to the $\overline{MS}$-scheme and 2-loop evolution from thereon one obtains the third entry in Table 3, in satisfactory agreement with the SF-value.

There are many papers which rely on global fitting procedures to extract the coupling. For example Klassen [18] proposes to obtain $\alpha_V$ by fitting the lattice data for the potential with a simple ansatz which assumes a particular form for the cutoff effects. The the expected continuum behavior in IR and UV is also built in and the associated parameters are obtained by optimizing the fits. However attractive a particular ansatz may be, the results depend sensitively on the form of the ansatz, and within such a procedure it is hard to make reliable error estimates.

### 5.3. The MOM-coupling

A preliminary study of the running of the MOM coupling in SU(3) Yang-Mills theory has recently appeared [20]. Unfortunately the momentum range covered at present is rather limited, but the authors claim their results are in satisfactory agreement with those obtained by the other methods.

## 6. USES OF BARE PT

At first it seems that the series eq.(11) is of little practical use since it only applies if $\alpha_0 \ln(qa)$ is small. But, as explained in [12], a similar situation is encountered in the continuum theory, when $\alpha_X(q)$ is expanded in powers of $\alpha_X(q')$ with $q' \gg q$. If we now choose $q'$ proportional to $1/a$ the series may be combined to give

$$\alpha_X(s/a) = \alpha_0 + d_1^X(s)\,\alpha_0^2 + d_2^X(s)\,\alpha_0^3 + \ldots. \quad (22)$$

It should be emphasized that in eq.(22) $\alpha_X(s/a)$ is defined in the continuum theory. In particular the momentum $s/a$ must be given in units of some physical reference scale (see sect.3) for the relation to become useful.

The formula above forms the basis of a more direct way to obtain the value of the running coupling at say the Z-peak pioneered by the group at FNAL [21,56]. The idea is 1) to measure a physical scale at a given bare coupling to get a value of $a(\alpha_0)$; 2) use a formula of the type of eq.(22) to get a value for $\alpha_{\overline{MS}}$ at $q = s/a$ and finally 3) run to $m_Z$ using the perturbative evolution.

The question is, what is the best way to proceed explicitly? If we choose for $N = 3$ e.g. $s = \Lambda_{\overline{MS}}/\Lambda_{\text{lat}} \sim 28.8$ (as suggested by previous experience), one gets $d_2^{\overline{MS}}(s) \sim 4.45$. In view of the size of this 2-loop correction and the large scale factor one may hesitate to make practical use of this form of the expansion. The problem with the scale factor disappears if instead of $\alpha_0$ one uses the "improved" bare coupling [42,56]

$$\tilde{\alpha}_0 = \alpha_0/P \quad (23)$$

as the expansion parameter [6]. Choosing again the scale factor so that the 1-loop term vanishes we end up with

$$\alpha_{\overline{\rm MS}}(2.63285/a) = \tilde\alpha_0 + 2.18505(1) \times \tilde\alpha_0^3 + \ldots . \quad (24)$$

This expansion looks more comfortable although in the relevant range of bare couplings, $g_0^2 \simeq 1$, the 2-loop correction is still 3–4%.

Lepage and Mackenzie [56] now propose using

$$\alpha_\Box = -p_1^{-1} \ln P. \quad (25)$$

Since the expansion of $P$ in $\alpha_0$ is known to 3-loops the corresponding expansion of $\alpha_\Box$ is known to 2-loops. e.g. for $N = 3, n_f = 0$ we have

$$\alpha_\Box = \alpha_0 + 3.373\alpha_0^2 + 17.70\alpha_0^3 + \ldots \quad (26)$$

At $\beta = 6.5$ this accounts for only 92% of the measured value, of which the last contribution gives 7%, thus this series seems poorly convergent.

However it is observed [56] that physical couplings have better perturbative series in parameters such as $\alpha_\Box$, which is of course also not a physical coupling but is expressed purely in terms of a measured quantity and as such may include some 'generic' non-perturbative effects. Going back for the moment to the case $N = 2$ one obtains

$$\alpha_{\rm SF}(a^{-1}) = \alpha_\Box + 0.731\alpha_\Box^2 + 1.125\alpha_\Box^3 + \ldots . \quad (27)$$

If one now uses the UKQCD value $\alpha_\Box = 0.1479(3)$ at $\beta = 2.85$, one gets $\alpha_{\rm SF}(20.6/r_0) = 0.1677(6)$, which compares favorably to the value $0.1693(23)$ which is obtained from the extrapolation of the Alpha-collaboration data to that point. Systematic deviations show up however when the procedure is repeated at smaller $\beta$.

Returning to $N = 3$ one has

$$\alpha_{\overline{\rm MS}}(q^*) = \alpha_\Box + 0.3632\alpha_\Box^2 + 1.017\alpha_\Box^3 + \ldots , \quad (28)$$

where the use of the scale $q^* = 3.41/a$, which is a sort of average value of $q$, $\ln q^{*2} = \overline{\ln q^2}$ occurring in the perturbative evaluation, is motivated in ref. [56]. The results (scale errors ommitted here) obtained from the above equation at $\beta = 6.5$ amputated at 1-loop and then at 2-loops (plus evolution to $q = 37/r_0$) are shown as the 4'th, 5'th entries in Table 3 respectively.

---

[6] This is because the dominant contributions to the 1-loop coefficient $d_1^{\overline{\rm MS}}(s)$ come from tadpole diagrams and that these cancel when $\alpha_0$ is replaced by $\tilde\alpha_0$

The NRQCD group [29] introduce $\alpha_{\rm P}$ through

$$\alpha_\Box = \alpha_{\rm P} - (1.185 + 0.070n_f)\alpha_{\rm P}^2, \quad (29)$$

where the coefficient multiplying $n_f$ also depends on the fermion action used (Wilson in this case). $\alpha_{\rm P}$ is not a physical coupling; but using measurements of $P$ and a physical scale at the same bare coupling one operationally assigns a value to $\alpha_{\rm P}(q^*)$ via eq.(29).

The coefficients in eq.(29) are chosen s.t.

$$\alpha_{\rm V}(q^*) = \alpha_{\rm P} + O(\alpha_{\rm P}^3). \quad (30)$$

To get to the $\overline{\rm MS}$-scheme one then often invokes

$$\alpha_{\overline{\rm MS}}(e^{-5/6}q) = \alpha_{\rm V}(q) + \frac{2N}{3\pi}\alpha_{\rm V}(q)^2 + \ldots, \quad (31)$$

where the scale is chosen so that the equation remains true for arbitrary $n_f$ [57].

Let us see how this prescription works out in the pure gauge theory. For example for SU(3) at $\beta = 6.0$ the scale $r_0/a = 5.44(26)$ [17]. Now use eq.(29) and the measurements of $\alpha_\Box$ to get an estimate for $\alpha_{\rm V}(q^*)$ and then use eq.(31) to get $\alpha_{\overline{\rm MS}}(e^{-5/6}q^*)$. Finally evolve via the RG with perturbative 3-loop $\beta$–function to the desired energy $q$. For $q = 37/r_0$ the result of this procedure is the 5'th entry in Table 3. Note only one error is shown corresponding mainly to the scale error. The second is an unknown systematic error because, one is working here at one value of the bare coupling, and hence an extrapolation to the continuum limit has not been made. The best we can do is to repeat the procedure for larger $\beta$ values and observe how the results converge. For example for $\beta = 6.4, 6.5$ one has $r_0/a = 9.90(54), 11.23(21)$ from [17], [15] resply., they give the same final result, the last entry in Table 3. Despite the fact that with this procedure a reliable error estimate is difficult to make, the results seem indeed in good agreement with those of SF.

## 7. EXTENSIONS TO FERMIONS

So far there has been no lattice investigation of the actual running of a coupling in QCD. But encouraged by the success of the FNAL method to get $\alpha_{\overline{\rm MS}}$ in pure gauge theories (see the previous section), many groups have been applying the method to fermions using quarkonium mass splittings to set the scale. Measurements have been made in the quenched approximation and for dynamical quarks with $n_f = 2$ with both Wilson and staggered fermion actions. To obtain values for $\alpha_{\overline{\mathrm{MS}}}$ in the realistic case there remain the principle problems of 1) the extrapolation to $n_f = 3$ and 2) the evolution over the heavy quark thresholds (a discussion of this is postponed to sect.9).

### 7.1. Quenched approximation

Many groups have investigated the quenched approximation: FNAL [21,22], NRQCD [28,29], UK(NR)QCD [32], KEK [31], CDHW [30] and ADHLM [33]. The groups all use some non-relativistic approximation, an expansion in powers of $v$ (or $1/m$), where $v$ is the velocity of the heavy quark in the quarkonium. However there are some differences in the the details of the approaches e.g. in the number of terms which are kept in the expansion, the way the light quarks are treated, and in some cases improved actions are used (see e.g. [33]) for the gauge theories and for the fermions. Despite these differences, which imply different cutoff effects, there is in general good agreement between the results of the groups for both charm and upsilon (see e.g. Figs.2 and 3 in ref.[25]). For a given onium the relative low-level splittings are in good agreement with experiment. For higher levels the agreement becomes a bit worse e.g. there is a 4-sigma difference in the measured value $(1P - 1S)/(2S - 1S) = 0.71$ in upsilon compared to the experimental value 0.78.

Compared to the situation reported by Michael last year [16], I am only aware of further progress by the FNAL [25] and NRQCD [27] groups, who have increased statistics and have concentrated on studying some of the systematic error sources. For example the FNAL group [24] found small cutoff effects in the $1P - 1S$ splittings, which translates to only a very slight $a$-dependence in $\alpha_{\overline{\mathrm{MS}}}^{(0)}(5\mathrm{GeV})$ determined from $b\bar{b}$ 1P-1S splitting.

Setting the scale, in the quenched approximation, using quantities with different characteristic scales $q^*$, one ends up with different estimates for the coupling e.g. there is a 5-6 sigma discrepancy in the NRQCD evaluation [27] of $\alpha_\mathrm{P}(8.2\mathrm{GeV})$ between upsilon and charm (see the first entries in

Table 4
Values of $\alpha_\mathrm{P}(8.2\mathrm{GeV})$ from NRQCD, $n_f = 3^*$ corresponds to extrapolation of $\alpha^{-1}$ to $n_f = 3$

| $n_f$ | scale | $\alpha_\mathrm{P}^{(n_f)}(8.2\mathrm{GeV})$ |
|---|---|---|
| 0 | cha. 1S-1P | 0.1480(13) |
| 0 | ups. 1S-1P | 0.1551(11) |
| 0 | ups. 1S-2S | 0.1505(09) |
| 2 | cha. 1S-1P | 0.1758(36) |
| 2 | ups. 1S-1P | 0.1793(11) |
| 2 | ups. 1S-2S | 0.1777(23) |
| $3^*$ | cha. 1S-1P | 0.1940(67) |
| $3^*$ | ups. 1S-1P | 0.1945(30) |
| $3^*$ | ups. 1S-2S | 0.1954(43) |

Table 4). This is however not surprising because the quenched approximation should not describe the running between the different $q^*$'s correctly, since $n_f$ is not the physical value.

A heuristic way which was used [21] to obtain an estimate for $n_f = 4$ from a quenched measurement, say in charmonium at $q^* \sim 5\mathrm{GeV}$, is to perturbatively run with $n_f = 0$ down to a typical charmonium scale $\sim 1\mathrm{GeV}$, and then integrate back up with $n_f = 4$. Within the quenched approximation alone it is however difficult to estimate the systematic errors involved in such a procedure.

In a recent review talk El-Khadra [7] summarized the following estimates for various errors contributing to the determination of $\alpha_{\overline{\mathrm{MS}}}^{(4)}(6 - 8\mathrm{GeV})$ in quenched calculations: statistical error $1 - 3\%$, finite $a$ $1 - 3\%$, perturbative 5%, sea quarks $5 - 8\%$. This gives a total of $\sim 8 - 10\%$ which at $m_Z$ amounts to $5 - 7\%$.

### 7.2. Dynamical quarks

The KEK group [31] were the first to report dynamical quark simulations measuring 1P-1S splittings and the plaquette expectation value $P$. They used $n_f = 2$ Wilson quarks and extrapolated $P$ to $m_q = 0$. They observed that integrating back the $n_f = 0, 2$ points, using the RG, the curves indeed come close at $E \sim 0.5\mathrm{GeV}$, which supports the original procedure adopted in ref.[21]. Encouraged by this they extended the method to go to $n_f = 3$ from their measurement at $n_f = 2$. Then evolving over the heavy quark thresholds and running to the Z-peak they obtained the result $\alpha_{\overline{\rm MS}}^{(5)}(m_Z) = 0.111(5)$.

---
[7]in Pittsburgh, June 1995

NRQCD has increased statistics on dynamical simulations compared to a year ago. A detailed report of their results can be found in these proceedings [27]. In particular they have studied the sensitivity of $\alpha_{\rm P}^{(n_f)}$ to: 1) the dynamical light quark mass $m_q$ and 2) on the physical quantity used to determine $a^{-1}$ (e.g. charmonium versus upsilon splittings). Concerning point 1) their new data at $am_q = 0.025$ is consistent with the statement that extrapolation from the bare mass value $am_q = 0.01$ used previously ($a^{-1} = 2.4$GeV) to realistic light quark masses has negligible effect compared to statistical errors.

Concerning point 2) they find that estimates of the coupling $\alpha_{\rm P}(8.2{\rm GeV})$ still depend on the physical quantity used to set the scale, but the differences are not as pronounced as in the quenched case. The estimates from various physical quantities have different dependences on $n_f$ and a comparison of $n_f = 0$ and $n_f = 2$ results suggests that $\alpha_P$ will become insensitive to which physical quantity is used to set the scale, once the correct number of dynamical flavors $n_f = 3$ is incorporated. Linearly extrapolating in $n_f$ their values of $\alpha_{\rm P}^{-1}$ at $n_f = 0, 2$ to $n_f = 3$, NRQCD [27] obtain the last three entries in Table 4 (see also Fig.5 of their proceedings contribution [27]), which are certainly consistent with one another. Such observations also explain most of last year's discrepancy between the central values of $\alpha_{\overline{\rm MS}}^{(5)} = 0.108(6)$ by Wingate et al [30] and $0.115(2)$ by NRQCD [28] who used charmonium respectively bottomium splittings to set the scale.

The measured value of $\alpha_{\rm P}$ has not changed during the past year, and NRQCD [27] still quote a final result of $\alpha_{\overline{\rm MS}}(m_Z) = 0.115(2)$, where the main error is considered as originating from the present ignorance of the 2-loop coefficient $d_2^{\overline{\rm MS}}$ for $n_f \neq 0$. If this happened to be close to its $n_f = 0$ value (of 0.96) then their central value would increase to 0.117.

## 8. THE $\alpha$-COLLABORATION PROJECT

The Alpha-collaboration has now completed detailed studies in pure gauge theories. Our program has so far been rather limited in physical extent; the stress has been more on precision (the control of systematic errors) and on the development of reliable methods applicable in more general situations.

We are now turning to QCD. The section at Rome II have been investigating "bermions" [58], a method first suggested in ref.[59] to take fermion effects approximately into account by extrapolation from negative $n_f$.

The main next goal is to measure the SF coupling in full QCD. The analytic framework is on a firm footing thanks to the work of Sint [60] who discussed the appropriate boundary conditions for the fermions and studied the renormalizability. Sint and Sommer [61] have completed various 1-loop computations of the SF-coupling including fermions; for $N = 3$ they obtained the term $\propto n_f$ in Table 1. They computed with Wilson fermions with and without the Sheikholeslami-Wohlert [62] improvement term, and showed that $O(a)$ cutoff terms at 1-loop could indeed be removed (with SW) by a modification of the weights of plaquettes touching the boundary. As a by-product they also obtained the ratio of the lattice $\Lambda$-parameters for the theories with and without improvement

$$\Lambda_{\rm SW}/\Lambda_{\rm W} = \exp\{0.02479675 n_f/(2b_0)\}. \qquad (32)$$

The program development for simulations on the QUADRICS (APE) is well under way for full QCD ($N = 3$). I would like to draw your attention to two useful studies which have arisen from the preparation of this project. The first by Jansen and Liu [63] who have made a study of the Kramer's algorithm first suggested by Horowitz [64]. The second is an efficient method for the computation of the eigenvalues of the lattice Dirac operator, by Kalkreuter and Simma [65]. Tests still remain to be made as to which fermion algorithm should be used. Recently there have been some improvements [66,67] to Lüscher's algorithm [68], so this looks rather promising.

The expansion of $\alpha_{\rm SF}$ along the lines explained in ref.[37] and the extension of the background field method for computing $\alpha_{\overline{\rm MS}}$ [8] in powers of $\alpha_0$ to 2-loop order in QCD with Wilson quarks is straightforward. Such computations would certainly be welcome, but it may take a while until the precision on $\alpha_{\rm SF}$ which can be reached in numerical simulations of full QCD is comparable to the magnitude of a typical 2-loop correction.

## 9. HEAVY QUARK THRESHOLDS

Among the first to consider how to take the heavy quark thresholds into account were Georgi and Politzer [69] already in 1975. Their considerations rested on the pioneering paper of Appelquist and Carazzone [70] in which their "decoupling theorem" was stated: *In a renormalizable field theory the heavy fields decouple at low momenta except for their contribution to renormalization effects.*

In the MS-scheme the $\beta$-function governing the running of the coupling $\bar{g}_{\rm MS}$ is independent of the quark mass. This is not in conflict with the theorem since when a Green function is expressed as a series in the MS-coupling, the terms of the series are such that they convert into a series in an effective coupling which behaves as if only the light degrees of freedom are present. The explicit demonstration of this to 2-loops is contained in a series of papers by Ovrut and Schnitzer [71] and by Wetzel [72]. Subsequently Bernreuther and Wetzel [73], assuming the AC theorem to hold for a momentum subtraction scheme, discussed the general relation (to all orders PT) between the couplings of the full theory and the effective light theory (valid in the momentum region $\ll m_{\rm heavy}$) in the MS-scheme.

How to take the heavy quark thresholds into account is thus quite well understood; a recent account has been given by Rodrigo and Santamaria [74]. The discussion in principle replaces ad hoc phenomenological matching relations such as $\bar{g}_{\rm MS}^{(4)}(cm_b) = \bar{g}_{\rm MS}^{(5)}(cm_b)$ (for some $c \sim 2$), but in practice such matching conditions yield results very close to those of a full analysis.

---

[8]This computation is being contemplated by R. Frezzotti

### 9.1. Mass dependence of the SF-coupling

The $\beta$-functions of all the couplings specified in sect.2, are quark mass $m$ dependent, and the running shows the decoupling of the heavy fermions. This can already be seen at 1-loop; the 1-loop coefficient takes the form,

$$b_0(n_f, z) = b_{0,0} + n_f b_{0,1}(z) \qquad (33)$$

where $z = m/q$ ($q$ the momentum with which the coupling runs). The mass dependence of $b_0(n_f, z)$ in the MOM-scheme was considered in [69], and that of the V-scheme is practically the same [75].

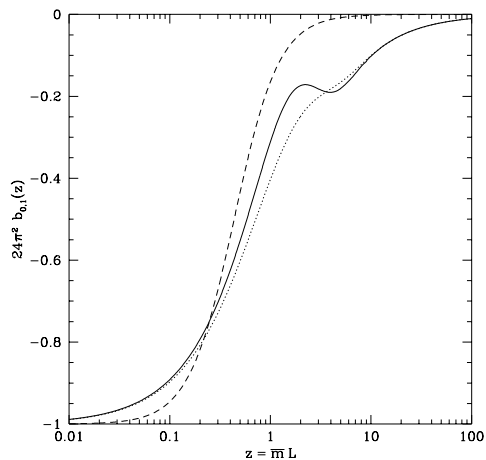

Figure 3. Contribution of one quark to the 1-loop $\beta$- function, for $\theta = 0$ (dotted line), $\theta = \pi/5$ (full line) for the V-scheme (dashed line). $\theta$ corresponds to different fermion spatial bc's.

In their recent paper Sint and Sommer [61] obtained $b_{0,1}(z)$ in the SF-scheme; it is plotted along with that of the V-scheme in Fig.3. The transition from an effectively massless quark to an approximately decoupled heavy quark in the SF-scheme is not as rapid as in the V-scheme. This is because in the SF-scheme there are corrections odd in $m$ ($\propto z$ as $z \to 0$, $\propto 1/z$ as $z \to \infty$), due to the 'temporal' fermion bc's. As discussed in ref.[61] the $1/z$ term in the $\beta$-function seems to pose a problem for the full non-perturbative computation along the lines of ref.[10], since quarks much heavier than $1/L$ induce lattice artifacts that dominate over their physical effect if one is limited to lattice sizes say $L/a < 20$. However one can in practice omit quarks with $z > z_{\text{cut}} \sim 2$, since if this is done then the error induced in $1/\bar{g}^2$ (at 1-loop) is only about 0.003 [61], which is a factor 10 less than the experimental error at LEP.

## 10. CONCLUSIONS

The various lattice determinations of $\alpha_{\overline{\text{MS}}}(m_Z)$ in pure gauge theory are consistent. These include treatments where extrapolation to the continuum limit is made carefully [6] -[13] and more direct approaches [21] where this is not the case.

The Alpha-collaboration have initiated a program to study the running coupling in full QCD. Until results of these (or analogous) studies are available, in order to make statements relevant to present phenomenology the "shortcut method" which worked well in the pure gauge case has been applied. This has been done for quenched [25] and for dynamical fermions. The $m_q, n_f, \beta$ dependence has been explored to some extent by the NRQCD group [27], but it still needs more work. It is crucial that the error sources are controlled and in particular systematic errors not underestimated. Probably a realistic goal in the near future is to match the experimental error $\leq 5\%$.

I thank all the members of the Alpha-collaboration for a very pleasant joint venture. I am very grateful to Aida El-Khadra for sending me useful information before my talk, and to Junko Shigemitsu for sending a summary of the recent NRQCD investigations. Finally I thank Chris Michael - much of this review was presented in his talk at Cortona in February this year.

**REFERENCES**

1. W. A. Bardeen, A. Buras, D. W. Duke and T. Muta, Phys. Rev. D18 (1978) 3998
2. see e.g. C. J. Morningstar, these proceedings
3. see e.g. C. A. Dominguez, Phys. Lett. B345 (1995) 291
4. B. R. Webber, Proc. Int. Conf. HEP, Glasgow 1994
5. see e.g P. H. Chankowski and S. Pokorski, Phys. Lett. B356 (1995) 307; hep-ph/9509207
6. M. Lüscher, P. Weisz and U. Wolff, Nucl. Phys. B359 (1991) 221
7. M. Lüscher, R. Narayanan, P. Weisz and U. Wolff, Nucl. Phys. B384 (1992) 168
8. M. Lüscher, R. Sommer, P. Weisz and U. Wolff, Nucl. Phys. B389 (1993) 247
9. M. Lüscher, R. Narayanan, R. Sommer, P. Weisz and U. Wolff, Nucl. Phys. B (Proc. Suppl.) 30 (1993) 139
10. M. Lüscher, R. Sommer, P. Weisz and U. Wolff, Nucl. Phys. B413 (1994) 481
11. G. de Divitiis, R. Frezzotti, M. Guagnelli and R. Petronzio, Nucl. Phys. B422 (1994) 382; Nucl. Phys. B433 (1995) 390
12. G. de Divitiis et al. (ALPHA collab.), Nucl. Phys. B437 (1995) 447
13. M. Guagnelli, Nucl. Phys. B (Proc. Suppl.) 42 (1995) 252
14. C. Michael, Phys. Lett. B283 (1992) 103
15. S. P. Booth, D. S. Henty, A. Hulsebos, A. C. Irving, C. Michael and P. W. Stephenson, Phys. Lett. B294 (1992) 385
16. C. Michael, Nucl. Phys. B (Proc. Suppl.) 42 (1995) 147
17. G. S. Bali and K. Schilling, Phys. Rev. D47 (1993) 661; Nucl. Phys. B (Proc. Suppl.) 34 (1994) 147
18. T. R. Klassen, Phys. Rev. D51 (1995) 5130 Nucl. Phys. B
19. M. Campostrini, P. Rossi and E. Vicari IFUP-TH 76/94
20. B. Allés, D. Henty, H. Panagopoulos, C. Parrinello and C. Pittori, LPTHE prep 95/24
21. A. X. El-Khadra, G. Hockney, A. S. Kronfeld and P. B. Mackenzie, Phys. Rev. Lett. 69 (1992) 729
22. A. X. El-Khadra, Nucl. Phys. B (Proc. Suppl.) 34 (1994) 141
23. G. P. Lepage and B. A. Thacker, Nucl. Phys. B (Proc. Suppl.) 4 (1988) 199; Phys. Rev. D43 (1991) 196
24. A. X. El-Khadra, G. Hockney, A. S. Kronfeld, P. B. Mackenzie, T. Onogi and J. Simone, Fermilab PUB-94/091-T
25. A. X. El-Khadra, hep-ph/9508266
26. G. P. Lepage, L. Magnea, C. Nakhleh, U. Magnea and K. Hornbostel, Phys. Rev. D46 (1992) 4052
27. P. McCallum and J. Shigemitsu, these proceedings ,

28. C. T. H. Davies, K. Hornbostel, A. Langnau, G. P. Lepage, A. Lidsey, J. Shigemitsu and J. Sloan, Phys. Rev. D50 (1994) 6963
29. C. T. H. Davies, K. Hornbostel, G. P. Lepage, A. Lidsey, J. Shigemitsu and J. Sloan, hep-lat/9506026; Phys. Rev. Lett. **73** (1994) 2654; Phys. Lett. B345 (1995) 42
30. M. Wingate, T. DeGrand, S. Collins, and U. Heller, Nucl. Phys. B (Proc. Suppl.) 34 (1994) 373; hep-lat/9501034
31. S. Aoki, M. Fukugita, S. Hashimoto, N. Ishizuka, H. Hino, M. Okawa, T. Onogi, A. Ukawa Phys. Rev. Letts. 75 (1995) 22
32. S. M. Catteral, F. R. Devlin, I. T. Drummond and R. R. Horgan, Phys. Lett. B300 (1993) 393; Phys. Lett. B321 (1994) 246
33. M. Alford, W. Dimm, G. Hockney, G. P. Lepage, P. B. Mackenzie, Nucl. Phys. B (Proc. Suppl) 42 (1995) 787; hep-lat/9507010
34. A. Patrascioiu and E. Seiler, AZPH-TH-92-06
35. K. Symanzik, Nucl. Phys. B190 [FS3] 1981 1
36. R. Sommer, Nucl. Phys. B411 (1994) 839
37. U. Wolff and R. Narayanan, Nucl. Phys. B (Proc. Suppl.) 42 (1995) 291; Nucl. Phys. B444 (1995) 425
38. M. Lüscher and P. Weisz, hep-lat/9505011; Phys. Letts. B349 (1995) 165
39. A. Hasenfratz and P. Hasenfratz, Phys. Lett. 93B (1980) 165; Nucl. Phys. B193 (1981) 210
40. R. K. Ellis and G. Martinelli, Phys. Lett. B141 (1984) 111; Nucl. Phys. B325 (1984) 93; (E: B249 (1985) 750)
41. R. K. Ellis, in: Proceedings of the Argonne National Laboratory Workshop (1984)
42. G. Parisi, in The Proceedings of the XX. Int. HE Conf. Madison (1980)
43. B. S. DeWitt, Phys. Rev. 162 (1967) 1195 and 1239
44. H. Kluberg-Stern and J. B. Zuber, Phys. Rev. D12 (1975) 482
45. R. Dashen and D. Gross, Phys. Rev. D23 (1981) 2340
46. M. Lüscher and P. Weisz, hep-lat/9504006
47. T. Reisz, Comm. Math. Phys. 116 (1988) 81 and 573; ibid 117 (1988) 79 and 639
48. M. Lüscher and P. Weisz, Nucl. Phys. B445 (1995) 429
49. A. Di Giacomo and G. C. Rossi, Phys. Lett. B100 (1981) 481
50. B. Allés, M. Campostrini, A. Feo and H. Panagopoulos, Phys. Lett. B326 (1994) 433
51. F. Di Renzo, G. Marchesini, P. Marenzoni and E. Onofri, Nucl. Phys. B (Proc. Suppl.) 42 (1995) 867
52. O. V. Tarasov, A. A. Vladimirov and A. Yu. Zharkov, Phys. Lett. 93B (1980) 429
53. C. Schlichter, G. S. Bali and K. Schilling, Nucl. Phys. B (Proc. Suppl.) 42 (1995) 273
54. M. Lüscher, Nucl. Phys. B180 [FS2] 1981 317
55. H. Wittig, Nucl. Phys. B (Proc. Suppl.) 42 (1995) 288
56. G. P. Lepage and P. B. Mackenzie, Phys. Rev. D48 (1993) 2250
57. S. J. Brodsky, G. P. Lepage and P. B. Mackenzie, Phys. Rev. D28 (1983) 228
58. G. M. de Divitiis, R. Frezzotti, M. Guagnelli, M. Masetti, R. Petronzio hep-lat/9507020
59. S. J. Anthony, C. Llewellyn-Smith and J. Wheater, Phys. Lett. 118B (1982) 287
60. S. Sint, Nucl. Phys. B421 (1994) 135; hep-lat/9504005
61. S. Sint and R. Sommer, hep-lat/9508012
62. B. Sheikholeslami and R. Wohlert, Nucl. Phys. B259 (1985) 572
63. S. Jansen and C. Liu, hep-lat/9506020
64. A. M. Horowitz, Phys. Lett. 156B (1985) 89; Nucl. Phys. B280 (1987) 510;
65. T. Kalkreuter and H. Simma, hep-lat/9507023
66. P. de Forcrand, these proceedings
67. B. Jegerlehner, in preparation
68. M. Lüscher, Nucl. Phys. B418 (1994) 637
69. H. Georgi and D. Politzer, Phys. Rev. D14 (1976) 1829
70. T. Appelquist and J. Carazzone, Phys. Rev. D11 (1975) 2856
71. B. A. Ovrut and H. J. Schnitzer, Phys. Rev. D22 (1980) 2518; Nucl. Phys. B179 (1981) 381; ibid B189 (1981) 509
72. W. Wetzel, Nucl. Phys. B196 (1982) 259
73. W. Bernreuther and W. Wetzel, Nucl. Phys. B197 (1982) 228
74. G. Rodrigo and A. Santamaria, Phys. Lett. B313 (1993) 441
75. S. Sint, private communication